\documentclass[11pt]{article}
\usepackage{times}
\usepackage{geometry}
\usepackage[utf8]{inputenc}
\usepackage{enumitem,amssymb}
\usepackage{ragged2e}
\usepackage{graphicx}
\usepackage{comment}
\usepackage{multicol}
\usepackage[usenames]{xcolor} %used for font color              
\definecolor{xlinkcolor}{cmyk}{1,1,0,0}
\usepackage{url}
\usepackage[
 colorlinks=true,    % false: boxed links; true: colored links 
 linkcolor=xlinkcolor,     % color of internal links            
 citecolor=xlinkcolor,     % color of links to bibliography
 filecolor=xlinkcolor,  % color of file links 
 urlcolor=xlinkcolor,      % color of external link
 final=true
]{hyperref}
\usepackage[super,sort&compress,comma]{natbib}
\usepackage{enumitem}
\setenumerate{itemsep=0mm}

\begin{document}
\begin{raggedright} 
% part of template, but does not look good
\huge
Snowmass2021 - Letter of Interest \hfill \\[+1em]
\textit{Determination of cosmic ray properties in the local interstellar medium with all-sky anisotropy observations} \hfill \\[+1em]
\end{raggedright}

\normalsize

\noindent {\large \bf Thematic Areas:}  %(check all that apply $\square$/$\blacksquare$)

%\noindent $\square$ (CF1) Dark Matter: Particle Like \\
%\noindent $\square$ (CF2) Dark Matter: Wavelike  \\ 
%\noindent $\square$ (CF3) Dark Matter: Cosmic Probes  \\
%\noindent $\square$ (CF4) Dark Energy and Cosmic Acceleration: The Modern Universe \\
%\noindent $\square$ (CF5) Dark Energy and Cosmic Acceleration: Cosmic Dawn and Before \\
%\noindent $\square$ (CF6) Dark Energy and Cosmic Acceleration: Complementarity of Probes and New Facilities \\
\noindent $\blacksquare$ (CF7) Cosmic Probes of Fundamental Physics \\
%\noindent $\square$ (Other) {\it [Please specify frontier/topical group]} \\

\noindent {\large \bf Contact Information:}\\
\begin{tabular}{ll}
Paolo Desiati$^1$           & [desiati@wipac.wisc.edu] \\
Juan Carlos Díaz Vélez$^1$  & [juancarlos@wipac.wisc.edu] \\
Nikolai Pogorelov$^2$       & [np0002@uah.edu] \\
Ming Zhang$^3$              & [mzhang@fit.edu] \\
\end{tabular}

\noindent {\large \bf Authors:} Paolo Desiati$^1$, Juan Carlos Díaz Vélez$^1$, Nikolai Pogorelov$^2$, Ming Zhang$^3$ \\[+1em]
{\small
$^1$ Wisconsin IceCube Particle Astrophysics Center (WIPAC), University of Wisconsin, Madison, U.S.A.\\
$^2$ University of Alabama in Huntsville, U.S.A.\\
$^3$ Florida Institute of Technology, U.S.A.\\
}

% ******************************************************************************************************
\noindent {\large \bf Abstract:}
Propagation of Galactic cosmic rays (CR) in the interstellar medium (ISM) is among the unsolved problems in particle astrophysics. Interpretation of CR spectrum and composition measurements and their possible link to dark matter crucially relies on our understanding of CR propagation in the Galaxy. Several air shower experiments have measured a significant anisotropy of CRs in the TeV to PeV energy range. These observations hint to a complicated overlap of more than one cause: from the distribution of the CR sources in the Milky Way to the nature of such sources, from the turbulence properties of interstellar plasmas to the inhomogeneous nature of the interstellar medium. Coherent magnetic structures such as the heliosphere greatly influence the CR arrival direction distribution. It is necessary to account for and remove the heliosphere's distortion effects if we want to determine the pristine CR arrival direction distribution in the local interstellar medium (LISM), the environment surrounding the solar system up to the distance of particle mean free path. The recent availability of accurate all-sky maps of CR arrival direction distribution and the latest advancements in heliospheric modeling, make it possible to infer the CR pitch angle distribution in the LISM using a Liouville mapping technique. With the interstellar CR distribution, we can study the global characteristics of CR diffusion, tap into the properties of interstellar plasma turbulence, test the recent and local CR source hypothesis, and whether clumps of dark matter have a role in the observed CR observations. The study can lead to developments aiming to a better understanding of the heliosphere, particularly the boundary region with the ISM, and additional constraints on the LISM properties.
% ******************************************************************************************************
\clearpage

% LOI text
%\noindent {\it Insert your white paper text here (maximum of 2 pages including figures).}
% ******************************************************************************************************
\noindent {\large \bf Introduction:}
An outstanding issue in astrophysics is the identification of the sources of CRs and how their energy spectrum and composition at Earth relates to that at the source. While gamma-ray and neutrino observations may provide hints into the remote CR injection spectrum, the observed CR flux at Earth is shaped by propagation in the ISM. The study of CR transport is the key to understanding their astrophysical origin.
So far, most investigations have relied on the information derived from CR energy spectrum measurements and composition. For example, the differences in power-law spectral slopes between the primary (H, He, C, etc.) and secondary (Li, Be, B, etc.) CR species offer a method of estimating the particle diffusion coefficient as a function of rigidity, which links CR spectra at the source to observations on Earth~\citep{strong1998, blasi2017}. However, recent observations with modern CR experiments have found that the energy spectra of several CR species (H, He, p$^-$, e$^-$, and e$^+$) can significantly deviate from a pure power-law, displaying bumps or valleys. No global CR propagation model can explain these features on the basis of spatially smoothed sources averaged over the CR residence time of many million years. Some researchers interpret the bumps as the contribution from one or a few local CR sources~\citep{blasi2009, mertsch2015, ahlers_2016}, while some others link them to an unknown interaction with dark matter particles~\citep{dev2014,huang2017}.
Another complication is that the diffusion coefficient is likely more complicated as it depends on the specific scattering properties of CR particles with magnetohydrodynamic (MHD) interstellar plasma~\citep{giacinti_kirk_2017, xu_laza_2020}. Accurate determination of CR pitch angle distribution in the ISM provides a direct probe of the interstellar turbulence and therefore more accurate diffusion properties.

Ground-based CR experiments can measure detailed maps of the CR flux's arrival direction distribution at TeV energy and higher. At ultra-high-energy scale (i.e., above EeV), CR particles point back to their extra-galactic sources with minimal deflections~\citep{auger_2017, TA_2018}. However, below the knee (i.e., around 3 PeV), CR particles are severely deflected and scattered by the interstellar magnetic field (ISMF) and its fluctuations. As a consequence, anisotropy can be used only to probe the overall diffusion-convection flow pattern of CRs. Diffusion may be the dominant cause of the observed anisotropy. In that case, the observations make it possible to explore the properties of particle scattering and density gradient in the LISM. A relatively close recent source may enhance the CR flux in a particular energy band of the energy spectrum observed at Earth. Time-dependent, local, individual contributions are highly sensitive to the particle diffusion coefficient, which is rigidity-dependent. For example, the anomaly of $^{22}$Ne/$^{20}$Ne ratio in GeV CRs may shed light onto contributions from supernovae of OB stars in the superbubble~\citep{binns2008}. In the meantime, the diffusive transport mechanism can leave its fingerprints on the CR anisotropy in the same energy band. Therefore, a combined study of the CR spectrum, composition, and anisotropy could provide us with a more conclusive validation of any theory describing the CR origin and transport. Anisotropy in e$^-$ and e$^+$ is considered vital information to distinguish if the e$^+$ bump in the spectrum comes from local supernova/pulsar sources or dark matter~\citep{adriani2015}. Furthermore, small-scale CR anisotropy is directly related to the ISMF turbulence~\citep{giacinti_sigl_2012, lb_2017}. Thus, it may provide crucial information on the turbulence spectrum and understanding of the driver of CR diffusion.

\noindent {\large \bf CR Anisotropy and the heliosphere :}
Several observations from large ground-based experiments have provided evidence of a small (up to order $10^{-3}$) but significant anisotropy of the CR flux at energies above several 10's GeV. This is especially true for the TeV-PeV energy range, which is covered by multiple experiments both in the northern and southern hemispheres~\citep{nagashima_1998, hall_1999, amenomori_2005, amenomori_2006, amenomori_2007, guillian_2007, abdo_2008, abdo_2009, aglietta_2009, zhang_2009, munakata_2010, amenomori_2011, dejong_2011, shuwang_2011, bartoli_2013, abeysekara_2014, bartoli_2015, amenomori_2017, bartoli_2018, abeysekara_2018, abbasi_2010, abbasi_2011, abbasi_2012, aartsen_2013, aartsen_2016}. With more data and substantial improvement in data analysis techniques, these observations reveal the CR anisotropy as a function of energy and angular scale with statistical accuracy in relative intensity below $10^{-5}$. However, the limited field of view of any individual ground-based experiment prevents us from capturing the anisotropy features at large angular scale. This limitation yields false results, as far as the properties of CR diffusion through the ISM are concerned. Combining observations from different ground-based observatories located in different hemispheres makes it possible to eliminate such limitations. The HAWC gamma-ray and the IceCube neutrino observatories have produced the first combined sky map of 10 TeV scale CR anisotropy~\citep{jcdv_2017}. In the future, with more and larger experiments, e.g., LHAASO~\citep{disciascio_2016} and the Southern Wide-field Gamma-ray Observatory (SWGO)~\citep{swgo_whitepaper, swgo_astro2020} being built or designed and by combining data from several experiments, it will be possible to obtain increasingly detailed anisotropy maps as a function of rigidity. All-sky unbiased CR arrival direction distributions provide a powerful tool to explore the origin of the observed CR anisotropy. In particular, they constitute a new tool to investigate of the heliospheric and interstellar magnetic fields~\citep{lazarian_desiati_2010, desiati_lazarian_2012, desiati_lazarian_2013, drury2008, drury2013, schwadron_2014, zhang_2014, lb_2016}. Beside a prominent dipole component seemingly aligned along the local ISMF, the observation maps of CR anisotropy shows additional significant contributions of medium and small scale features down to a few degrees~\citep{jcdv_2017, icrc2019}. Several features show correlations with heliospheric effects. For instance, the flux enhancement referred to as region A, lies along the heliosphere's tail direction, and in association with the location of the $B$-$V$ plane (the plane formed by the interstellar velocity and magnetic field directions deep in the LISM). With the availability of these new CR observations, it is possible at last to account for the warping effects of the heliosphere on 1-100 TeV scale CR flux, and unfold their gradient density and pitch angle distribution in the ISM. Ultimately, this will help us reveal the physics of CR diffusion in the ISM and the turbulence affecting it.

The possibility of unfolding the effects of the heliosphere from the observed CR arrival direction distribution constitutes a new drive to develop novel detailed heliospheric models with the emphasis on the solar wind-ISM boundary. Recent modeling involves adaptive mesh refinement numerical integration of MHD equations for plasma coupled with multi-fluid kinetic transport for neutral atoms~\citep{Pogo14}. This model, which accounts for all plasma/magnetic field and neutral gas interactions, was originally developed to make predictions for~\emph{Voyager} interstellar mission and interpret Interstellar Boundary EXplorer (IBEX) observations~\citep{McComas09, Jacob10, Erik16}. Now both \emph{Voyager} 1 and 2 are in the LISM, making in-situ measurements of the local interstellar magnetic field and plasma properties. Observations just outside the heliosphere~\citep{Stone05,Stone08,Stone13} can greatly constrain the LISM parameters. The model has been validated against numerous in-situ and remote observations, e.g., (\emph{SOHO} Ly$\alpha$ back-scattered emission, Ly$\alpha$ absorption profiles in directions towards nearby stars, \textit{New Horizons} observations in the distant SW, in-situ measurements in the SW and LISM from \textit{Voyagers}, etc.)~\citep{Kim16, Kim17, Pogo17a, Pogo17b}. Although we do not expect the LISM conditions to change within decades of CR measurements, the most recent models take into account solar cycle effects with the input of remote measurements of the photospheric magnetic field and initiate coronal mass ejections using multi-viewpoint observations \citep{Pogo16, Yalim17, Talwinder18, Talwinder19}. In this way, it is possible to investigate the potential minor time-dependence of CR anisotropy. Since TeV CR are sensitive to the transverse size of the heliosphere, in particular to the draping of the local ISMF fieldlines around the flanks, the study of CR flux in the LISM may provide invaluable hints into the interstellar heliospheric boundary region.

\noindent {\large \bf Heliospheric distortion of CR flux:}
Due to the significant influence that the heliosphere has on the CR arrival direction distribution up to a few 100s TeV scale, it is necessary to subtract those effects from the observations if we want to know what the CR flux looks like in the ISM itself. Accurate and unbiased all-sky maps of the CR flux, along with state-of-the-art modeling of the heliosphere, are necessary to reach such goal. Advanced CR observations and modeling capabilities have matured enough to make it possible to perform meaningful studies on the CR origin and propagation. Using a Liouville mapping technique, which takes into account the detailed heliospheric magnetic field structure, is employed. With such a method, it is possible to derive the density gradient and pitch angle distribution of TeV CRs in the LISM while accounting for particle trajectory chaotic behavior~\citep{vanessa_icrc2019}, and the residual experimental systematic biases~\citep{icrc2019} affecting the CR anisotropy sky maps. With such results, we will be able to probe the global CR propagation through the ISM. Future refinements will benefit from additional improvements in the heliospheric modeling and experimental determination of the anisotropy for different CR species over a wide energy range. Similar studies may be done with e$^{-}$e$^{+}$ anisotropy, once the observations have enough accurate statistical determination. The determination of hadronic and leptonic CR distributions beyond the heliosphere's influence will prove a powerful tool to the origin of the CRs. It may provide hints of nearby recent sources or indirect evidence of Dark Matter clumps in the ISM.

\clearpage
%\noindent {\large \bf References:} \\

\vspace{3.5in}

%\noindent {\large \bf Additional Authors:}\\

\end{document}